\documentclass[pdflatex, sn-basic, Numbered]{sn-jnl}

\usepackage{graphicx}%
\usepackage{multirow}%
\usepackage{amsmath,amssymb,amsfonts}%
\usepackage{amsthm}%
\usepackage{mathrsfs}%
\usepackage[title]{appendix}%
\usepackage{xcolor}%
\usepackage{textcomp}%
\usepackage{manyfoot}%
\usepackage{booktabs}%
\usepackage{algorithm}%
\usepackage{algorithmicx}%
\usepackage{algpseudocode}%
\usepackage{listings}%
\usepackage{longtable}
\usepackage{adjustbox}
\usepackage{caption}
\usepackage{hyperref}

\theoremstyle{thmstyleone}%
\theoremstyle{thmstyletwo}%

\theoremstyle{thmstylethree}%

\begin{document}

\title[Article Title]{Mexico 2021: Psychological Intimate Partner Violence Against Women and the Role of Childhood Violence Exposure - A Machine Learning Approach}


\author*[1]{\fnm{Clara} \sur{Strasser Ceballos }}\email{clara.strasserceballos@stat.uni-muenchen.de}

\author[1,2]{\fnm{Anna-Carolina} \sur{Haensch }}\email{anna-carolina.haensch@stat.uni-muenchen.de}

\affil*[1]{\orgdiv{Institute of Statistics}, \orgname{Ludwig-Maximilian University of Munich}, \orgaddress{\street{Ludwigstr. 33}, \city{Munich}, \postcode{80539}, \state{Bavaria}, \country{Germany}}}

\affil[2]{\orgdiv{Joint Program in Survey Methodology}, \orgname{University of Maryland}, \orgaddress{\street{7251 Preinkert Dr}, \city{College Park}, \postcode{20742}, \state{MD}, \country{United States}}}


\abstract{In 2021, psychological violence was the most prevalent form of intimate partner violence (IPV) suffered by women in Mexico. The consequences of psychological IPV can include low self-esteem, depression, and even potential suicide. It is, therefore, crucial to identify the most relevant risk and protective factors of psychological IPV against women in Mexico. To this end, we adopt an ecological approach and analyze the role of a wide range of factors across four interrelated levels: Individual, relationship, community, and societal. We construct a multidimensional data set with 61,205 observations and 59 variables by integrating nationally representative data from the 2021 Mexican Survey on the Dynamics of Household Relationships with nine additional sources. For model estimation and factor selection, we combine model-based boosting with stability selection. Our findings reveal that women who were exposed to violence in childhood and whose partners were exposed to violence in childhood face a heightened risk of psychological IPV. These findings highlight the critical yet often overlooked role of childhood violence exposure for psychological IPV risk. Additionally, we confirm the role of three protective factors previously identified by \citet{torresmunguiaDeterminantsEmotionalIntimate2022} using 2016 data, now validated with 2021 data: women who had their first sex later in life and under consent, who have autonomy in decision-making regarding their professional life and use of economic resources, and who live in a household where housework is done only by male members face a lower risk of suffering psychological IPV.}




\maketitle

\section*{Acknowledgments}
We want to thank Juan Armando Torres Munguía and Inmaculada Martínez-Zarzoso for providing us with the replication code and for answering our questions during the research.

\section{Introduction}\label{sec1}

In 2021, according to the nationally representative Survey on the Dynamics of Household Relationships in Mexico (ENDIREH 2021), 23.2\% of women aged 15 or older reported having experienced psychological intimate partner violence (IPV) in their current relationship \citep{inegiENDIREHMainResults2021}. Psychological IPV encompasses a range of harmful behaviors, including jealousy, intimidation, stalking, marginalization, and threats of physical violence, all of which can have severe consequences for women's mental and physical health \citep{inegiENDIREHMainResults2021}. The alarmingly high prevalence of psychological IPV and its potential consequences highlight the urgent need for effective interventions. Against this background, we aim to quantitatively identify both the groups of women most at risk of experiencing psychological IPV and the factors that may help mitigate this risk.

We draw on the approach presented by \citet{torresmunguiaDeterminantsEmotionalIntimate2022} and adopt the following methodological strategy. We use an ecological model, i.e., we consider a set of theoretically motivated factors and their interactions at four interrelated levels: Individual, relationship, community, and societal \citep{cdcViolencePrevention2024}.  To construct these factors, we draw on a new data source, the most recent Survey on the Dynamics of Household Relationships in Mexico for 2021 (ENDIREH 2021) \citep{inegiEncuestaNacionalSobre2021}. Further, we use nine additional data sources to include community and societal-level factors neglected in existing studies \citep{krugWorldReportViolence2002}. Using these 10 data sources allows us to construct a multidimensional data set with 61,205 observations and 59 variables. We use model-based boosting to estimate the relationship between these variables and women's psychological IPV risk. In addition to effect estimation, this algorithm offers two important features: automated variable selection and model choice during the fitting process \citep{hofnerModelbasedBoostingHandson2014}. Hence, the algorithm identifies the most relevant variables and chooses their appropriate
modeling alternative, e.g., whether the relationship is linear, non-linear, or both \citep{kneibVariableSelectionModel2009}. Further, we use the stability selection algorithm to narrow the number of selected variables to the most significant ones \citep{hofnerControllingFalseDiscoveries2015}.

Our contribution is threefold. First, we provide an extensive overview of the studies that have analyzed psychological IPV in Mexico. Second, we validate the work by \citet{torresmunguiaDeterminantsEmotionalIntimate2022} using new data from 2021 instead of 2016. Finally, we incorporate several previously understudied factors at the individual and relationship levels for psychological IPV \citep{guedesBridgingGapsGlobal2016}. Specifically, we consider the role of childhood violence exposure, which was not considered by \citet{torresmunguiaDeterminantsEmotionalIntimate2022}, the only other study employing an extensive ecological
approach to IPV in Mexico.

Our results show that women who experienced or witnessed psychological or physical violence in their childhood, who suffered sexual violence during their childhood, or who have partners who were exposed to violence in childhood face a heightened risk of psychological IPV. Additionally, we confirm several findings from \citet{torresmunguiaDeterminantsEmotionalIntimate2022}. Specifically, women with consensual late-life first-sex experiences, those living in households where the household chores are done exclusively by male household members, and those with moderate decision-making power concerning professional and economic resource decisions are less vulnerable to psychological IPV. Moreover, women residing in states with high crime rates against men are at higher risk.

The paper is structured as follows. First, we review existing literature (Section \ref{lit_review}). Second, we outline the methodology, including the data used and our estimation approach (Section \ref{method}). Further, we present and discuss the results of our primary analysis (Section \ref{results}). Finally, we close with a summary and future research suggestions (Section \ref{conclusion}).

\section{Literature Review}
\label{lit_review}

Identifying the determinants of IPV against women requires a thorough understanding of prior research findings. However, direct comparisons are challenging due to substantial variations in research approaches. Studies on IPV against women in Mexico differ across seven key dimensions, making direct comparisons complex: First, the studies differ in terms of their spatial focus (national, e.g., \citet{bottCorrelatesCooccurringPhysical2022} vs. regional, e.g.,
\citet{lopezRiskFactorsIntimate2015} and \citet{aguerrebereIntimatePartnerViolence2021}). Second, they differ in their temporal focus (2006,
e.g., \citet{terrazas-carrilloEmploymentStatusIntimate2015} vs. 2016, e.g., \citet{torresmunguiaDeterminantsEmotionalIntimate2022}). Third, they vary in the use of data sources for analysis (national surveys, e.g., \citet{avila-burgosFactorsAssociatedSeverity2009} vs. own survey instruments, e.g., \citet{valdez-santiagoPrevalenceSeverityIntimate2013}). Further, they vary
in the specific type of IPV they investigate (psychological, e.g., \citet{torresmunguiaDeterminantsEmotionalIntimate2022} vs. economical, e.g., \citet{casiqueCambiosConstantesNiveles2019} vs. physical, e.g., \citet{castroEmpowermentPhysicalViolence2008} vs. sexual, e.g., \citet{canedoEstimationEffectWomen2021}). Fifth, they differ with regard to the groups
of women for whom risk factors are analyzed (women with children, e.g., \citet{torresmunguiaDeterminantsEmotionalIntimate2022} vs. dating women between 15-24 years, e.g., \citet{rodriguez-hernandezPrevalenciaCorrelatosViolencia2023} vs. women older than 64 years, e.g., \citet{hernandezPrevalenciaFactoresAsociados2020}).
Sixth, there are, in general, differences in the selection and inclusion of factors, which are predominantly determined by data
availability and spatiotemporal focus of analysis (only individual and relationship-level
factors, e.g., \citet{lopezRiskFactorsIntimate2015} vs. additional municipality and state-level factors, e.g.,
\citet{valdez-santiagoPrevalenceSeverityIntimate2013}). Finally, they differ in the methodological approach used to
find the significant factors (logit model e.g., \citet{casiqueCambiosConstantesNiveles2019} vs. machine learning,
e.g., \citet{torresmunguiaDeterminantsEmotionalIntimate2022} vs. propensity score matching, e.g., \citet{canedoEstimationEffectWomen2021} vs. fixed-effect strategy, e.g., \citet{silverio-murilloRemittancesDomesticViolence2022}
vs. Difference-in-Difference (DiD), e.g., \citet{garcia-ramosDivorceLawsIntimate2021}).

As shown in Table \ref{tab_lit_rev}, most studies on psychological IPV focus on individual- and relationship-level factors, while community- and societal-level factors remain under-explored. This gap likely stems from the limited inclusion of such variables in past analyses, with \citet{torresmunguiaDeterminantsEmotionalIntimate2022} being a notable exception. To provide a structured overview, Table \ref{tab_lit_rev} summarizes key findings, where a negative sign indicates a protective factor and a positive sign denotes a risk factor.

\begin{longtable}{p{7cm}|p{4cm}|p{0.8cm}}
    \caption{Significant factors across four levels. The factors correspond to significant levels of *p$<$.10. **p$<$.05. ***p$<$.01. In \citet{torresmunguiaDeterminantsEmotionalIntimate2022}, they correspond to factors selected by the stability selection algorithm \citep{hofnerControllingFalseDiscoveries2015}. } 
    \label{tab_lit_rev} \\
\hline
\textbf{Factors} & \textbf{Previous Research} & \textbf{Effect} \\
\hline
\endfirsthead

\hline
\textbf{Factors} & \textbf{Previous Research} & \textbf{Effect} \\
\hline
\endhead

\hline
\endfoot

\hline
\endlastfoot

\multicolumn{3}{l}{\textbf{Individual-Level}} \\ \hline
Woman’s age in years & \citet{casiqueFactoresEmpoderamientoProteccion2010} & - \\
& \citet{casiqueCambiosConstantesNiveles2019} & - \\
 &  \citet{torresmunguiaDeterminantsEmotionalIntimate2022} & - \\ \hline
Woman’s age at first-sex by condition of consent yes (ref: no) & \citet{torresmunguiaDeterminantsEmotionalIntimate2022} & - \\ \hline
Woman has a disability yes (ref: no) & \citet{rodriguez-hernandezPrevalenciaCorrelatosViolencia2023} & + \\ \hline
Woman attended school yes (ref: no) & \citet{casiqueFactoresEmpoderamientoProteccion2010} & + \\
& \citet{casiqueCambiosConstantesNiveles2019} & + \\ \hline
Woman receives remittances yes (ref: no) & \citet{silverio-murilloRemittancesDomesticViolence2022} & + \\ \hline
Woman is employed yes (ref: no) & \citet{casiqueCambiosConstantesNiveles2019} & + \\
 &  \citet{rodriguez-hernandezPrevalenciaCorrelatosViolencia2023}  & + \\ \hline
Relationship status is cohabitation (ref: married) & \citet{casiqueFactoresEmpoderamientoProteccion2010} & + \\
 & \citet{casiqueCambiosConstantesNiveles2019} & + \\ \hline
Number of previous relationships & \citet{casiqueCambiosConstantesNiveles2019}  & + \\
 & \citet{rodriguez-hernandezPrevalenciaCorrelatosViolencia2023}  & + \\ \hline
Time in the relationship & \citet{rodriguez-hernandezPrevalenciaCorrelatosViolencia2023}  & + \\ \hline
Woman’s age at marriage or cohabitation with current partner & \citet{casiqueCambiosConstantesNiveles2019} & - \\ \hline
Woman’s age at cohabitation or marriage with current partner by condition of consent yes (ref: no) & \citet{torresmunguiaDeterminantsEmotionalIntimate2022} & + \\ \hline
Violence witnessed in childhood by woman yes (ref: no) & \citet{casiqueFactoresEmpoderamientoProteccion2010} & + \\
& \citet{casiqueCambiosConstantesNiveles2019} & + \\ \hline
Violence experienced in childhood by woman yes (ref: no) & \citet{casiqueFactoresEmpoderamientoProteccion2010} & + \\
 & \citet{casiqueCambiosConstantesNiveles2019} & + \\
 &  \citet{rodriguez-hernandezPrevalenciaCorrelatosViolencia2023}  & + \\ \hline
Number of children & \citet{casiqueFactoresEmpoderamientoProteccion2010} & + \\
& \citet{aguerrebereIntimatePartnerViolence2021} & + \\ \hline
Woman’s decision-making power (index) & \citet{casiqueCambiosConstantesNiveles2019} & - \\ \hline
Pro-gender equality attitude of woman (index) & \citet{casiqueFactoresEmpoderamientoProteccion2010} & + \\
 & \citet{casiqueCambiosConstantesNiveles2019} & + \\
  & \citet{rodriguez-hernandezPrevalenciaCorrelatosViolencia2023}  & - \\
\hline
\multicolumn{3}{l}{\textbf{Relationship-Level}} \\ \hline
Partner is 5 or more years older than woman & \citet{casiqueCambiosConstantesNiveles2019} & - \\ \hline
Woman is 2 or 3 years older than partner & \citet{casiqueCambiosConstantesNiveles2019} & + \\ \hline
Both woman and partner speak an indigenous language yes (ref: no one speaks) & \citet{casiqueCambiosConstantesNiveles2019} & - \\ \hline
Partner has children from previous relationships yes (ref: no) & \citet{casiqueCambiosConstantesNiveles2019} & + \\ \hline
Violence witnessed in childhood by partner yes (ref: no) & \citet{casiqueFactoresEmpoderamientoProteccion2010} & + \\ \hline
Violence experienced in childhood by partner yes (ref: no) & \citet{casiqueFactoresEmpoderamientoProteccion2010} & + \\
 &  \citet{casiqueCambiosConstantesNiveles2019} & + \\ \hline
Alcohol and drug use of partner yes (ref: no) & \citet{aguerrebereIntimatePartnerViolence2021} & + \\ \hline
Woman's participation in domestic work (index) & \citet{casiqueFactoresEmpoderamientoProteccion2010} & + \\ 
& \citet{casiqueCambiosConstantesNiveles2019} & + \\ \hline
Division of housework only among male members (ref: only female members) & \citet{torresmunguiaDeterminantsEmotionalIntimate2022} & - \\ \hline
Woman can decide when to have sex yes (ref: no) & \citet{casiqueFactoresEmpoderamientoProteccion2010} & - \\ \hline
Woman’s autonomy about professional life and use of economic resources medium level (ref: low) & \citet{torresmunguiaDeterminantsEmotionalIntimate2022} & - \\ \hline
Woman attends religious gatherings yes (ref: no) & \citet{casiqueFactoresEmpoderamientoProteccion2010} & - \\ \hline
Woman’s perception about support from social network medium level (ref: low) & \citet{torresmunguiaDeterminantsEmotionalIntimate2022} & + \\ \hline

\multicolumn{3}{l}{\textbf{Community-Level}} \\ \hline
Municipality type is urban (ref: rural) & \citet{casiqueFactoresEmpoderamientoProteccion2010} & + \\
 & \citet{casiqueCambiosConstantesNiveles2019}  & + \\
 &  \citet{rodriguez-hernandezPrevalenciaCorrelatosViolencia2023}  & + \\ \hline
Gini index & \citet{torresmunguiaDeterminantsEmotionalIntimate2022} & Inverted U-Shape \\ \hline
Women’s economically active population in the municipality & \citet{torresmunguiaDeterminantsEmotionalIntimate2022} & + \\ \hline
\multicolumn{3}{l}{\textbf{Societal-Level}} \\ \hline
Unilateral divorce laws & García-Ramos, 2021 & + \\ \hline
Prevalence of common crimes against men in federal state & \citet{torresmunguiaDeterminantsEmotionalIntimate2022} & + \\ 

\end{longtable}



\section{Methodology}
\label{method}
Our analysis focuses on psychological IPV against women with at least one child at the national level in 2021. We use national surveys and administrative records as data sources.
Our methodological approach to identify the most relevant risk and protective factors follows \citet{torresmunguiaDeterminantsEmotionalIntimate2022} and consists of four key stages. First, we define a set of theoretically motivated factors for each level of the ecological model. We then collect data on these factors from a variety of sources. Third, we specify our model. Finally, we apply model-based boosting and stability selection to identify the most significant factors.

\subsection{Data}
\subsubsection{Main Data Source}
At the heart of our study lies the the ENDIREH 2021. The survey is conducted by the National Institute of Statistics and Geography (INEGI) \citep{inegiEncuestaNacionalSobre2021}. The ENDIREH 2021 is preceded by four earlier publications: ENDIREH 2003, 2011, 2013 and 2016 \citep{inegiENDIREHInformeOperativo2021}. The survey collects information on the psychological, economic, physical, and sexual violence experienced by women aged 15 and over during their lifetime and in the last year (October 2020 – October 2021) in the family, school, workplace, community, and intimate partnership in Mexico \citep{inegiENDIREHMainResults2021}. The sample design of the survey has the following characteristics: probabilistic, three-stage, stratified, and clustered sampling 
\citep{inegiENDIREHDisenoMuestral2021}. Hence, the design ensures sample validity and guarantees nationally and federally representative information. At the national level, information about women’s experiences of violence was collected for 110,127 women aged 15 and over \citep{inegiENDIREHDisenoMuestral2021}. In our study, we use this survey data set to get information on the psychological violence experienced by women aged 15 and over with at least one child within the intimate partnership between 2020 and 2021. Further, we use the data set to construct the ecological model's pre-specified individual and relationship-level factors. More information on the survey data set can be found in Appendix \ref{app_survey_data}.

\subsubsection{Further Data Sources}
The ENDIREH 2021 survey further collects information on women’s municipality and federal state of residence. Hence, this allows us to create community and societal-level factors from nine additional data sources and link them to the individual observations of the survey data set. To construct community-level factors, we use the following municipality-level data: homicide records \citep{inegiMortalidad2021} and
migration data \citep{inegiMigracion2021} collected 2021 by INEGI, data from the 2020 Population and Housing Census (Intercensal) \citep{inegiCensoPoblacionVivienda2020}, the constructed Gini Index from the 2020 National Council for the Evaluation of Social Development Policy (CONEVAL) \citep{conevalMedicionPobreza2020}, the constructed Human Development Index from the 2020 United Nations Development Program  (UNDP) \citep{undpHumanDevelopmentIndex2020}, data from the 2020 National Census of Municipal and Delegation Governments (CNGMD) \citep{inegiCensoNacionalGobiernos2020}, and marginalization data from the National Population Council (CONAPO) \citep{conapoIndicesMarginacion2020}. To obtain state-level information, we use crime data from the 2021 National Survey on Victimization and Perception of Public Safety (ENVIPE) 
\citep{inegiEncuestaNacionalVictimizacion2021} and population satisfaction data from the 2019 National Survey of Quality and Governmental Impact (ENCIG) \citep{inegiEncuestaNacionalCalidad2019}.

\subsubsection{Final Data Set}
We use all ten data sources to construct our final data set, with the ENDIREH 2021 as the core. To this end, we conduct a rigorous data preparation process consisting of five main steps. First, we filter the ENDIREH 2021 to women aged 15 and above, married or cohabiting with a male partner, and having at least one child. Second, we link information from the additional nine data sources at the state and municipality levels to the filtered data set. Third, we use predictive mean matching to impute missing data and obtain a data set with complete cases \citep{buurenFlexibleImputationMissing2018}. Fourth, we conduct a thorough review to identify and eliminate implausible observations within the data set. Finally, we remove outliers. The final data set contains 61,205 observations. A detailed overview of the data preparation steps can be found in Appendix \ref{app_data_prep}.

\subsection{Variables}

\subsubsection{Outcome Variable }
The outcome variable takes the value 1 if a woman suffered psychological IPV in the last year (October 2020 to October 2021) and 0 otherwise. Questionnaire A, the questionnaire for married or cohabiting women, of the ENDIREH 2021 comprises 15 questions that are specifically designed to determine whether women have encountered instances of psychological IPV within the past year or not \citep{inegiEncuestaNacionalSobre2021}. The initial phrase introducing the set of 15 questions is formulated as, ``From October 2020 to date, has this occurred…''. Subsequent questions include specific incidents, such as ``… made you feel afraid?.'' Women are provided with response options including ``often,'' ``sometimes,'' ``once,'' and ``it did not occur.'' The resulting outcome variable is assigned a value of 1 if the woman responded ``once'' to at least one of the 15 questions and a value of 0 if the woman indicated ``it did not occur'' for all the questions. Table \ref{tab:emo_ipv} presents the corresponding behaviors from the ENDIREH 2021 for the different psychological IPV act types. 

\begin{table}[h!]
\caption{Psychological IPV acts and behaviors. The table is based on \citet{torresmunguiaDeterminantsEmotionalIntimate2022}, with minor changes}
\label{tab:emo_ipv}
    \small 
\begin{tabular}{|p{0.2\textwidth}|p{0.7\textwidth}|}
\hline
    \textbf{Act Type} & \textbf{Behaviour Description} \\ \hline
     Indifference & $\cdot$ Ignoring, disregarding, or lacking affection towards the woman. \\
     & $\cdot$ Stop talking to the woman. \\
     \hline
    Insults, Humiliation, and Devaluation & $\cdot$ Embarrassing, insulting, degrading, or humiliating the woman (calling her ugly or comparing her to other women). \\
    \hline
     Jealousy & $\cdot$ Accusing the woman of cheating on him. \\
     \hline
    Intimidation & $\cdot$ Making the woman feel afraid. \\
     & $\cdot$ Destroying, throwing away, or hiding objects belonging to the woman or the household. \\
      & $\cdot$ Being angry with the woman because the chores are not done, because the food does not meet his expectations, or he feels that the woman has not fulfilled her obligations. \\
      \hline
       Stalking & $\cdot$ Watching the woman, spying on her, following her, showing up to her unexpectedly. \\
     & $\cdot$ Calling or texting the woman repeatedly to find out where she is, if she is with someone, and what she is doing. \\
     & $\cdot$ Monitoring the woman's e-mails or mobile phone and demanding her passwords. \\
      \hline
    Marginalization & $\cdot$ Forbidding the woman to leave the house, locking her up, or stopping her from having visits. \\
     & $\cdot$ Turning the woman's children or relatives against the woman. \\
    \hline
    Threats & $\cdot$ Threatening to leave/abandon the woman, to harm her, to take away her children, or to kick her out of the house. \\
    & $\cdot$ Threatening the woman with a weapon (knife, blade, gun, or rifle) or with burning her. \\
     & $\cdot$ Threatening to kill the woman, to kill himself, or to kill the children. \\
     \hline
\end{tabular}
\end{table}

\subsubsection{Factors}

We create 59 theoretically motivated factors from the information provided by the ten data sources (see Table \ref{tab:risk_factors}). New variables that have not been previously examined by \citet{torresmunguiaDeterminantsEmotionalIntimate2022} include the woman’s occupational status, whether she receives conditional cash transfers (CCT), whether she has had previous cohabitation or marriage experiences, whether she has witnessed psychological or physical violence during childhood, whether she has experienced psychological or physical violence during childhood, whether she has experienced sexual violence during childhood, the number of children, whether the partner has witnessed psychological or physical violence during childhood, and whether the partner has experienced such violence during childhood. A detailed variable codebook is included in Appendix \ref{app_code_book}.

\begin{center}
    \small 
    \begin{longtable}{|p{0.7\textwidth}|p{0.2\textwidth}|}
    \caption{Factors and data source}
    \label{tab:risk_factors}
     \\ \hline
    \textbf{Factor (Type)} & \textbf{Source} \\
    \hline
    \endfirsthead
    \multicolumn{2}{l}{\textit{}} \\
    \hline
    \textbf{Factor (Type)} & \textbf{Source} \\
    \hline
    \endhead
    \hline
    \multicolumn{2}{r}{\textit{}} \\
    \endfoot
    \hline
    \endlastfoot
\multicolumn{2}{|l|}{\textbf{Individual-Level}} \\
\hline
Age of woman in years (cont.) & ENDIREH 2021\\
Education level of woman (cat.: low, medium, high) & ENDIREH 2021\\
Indigenous woman (cat.: no, yes) & ENDIREH 2021\\
Income per month of woman (cont.) & ENDIREH 2021\\
CCT receiver (cat.: no, yes) & ENDIREH 2021\\
Unemployment in the last 12 months (cat.: no, yes) & ENDIREH 2021\\
Previous cohabitation or marriage (cat.: no, yes) & ENDIREH 2021\\
Violence witnessed in childhood (cat.: no, yes) & ENDIREH 2021\\
Violence experienced in childhood (cat.: no, yes) & ENDIREH 2021\\
Sexual violence experienced in childhood (cat.: no, yes) & ENDIREH 2021\\
Number of children (cont.) & ENDIREH 2021\\
Age at first child (cont.) & ENDIREH 2021\\
Age at first sexual intercourse (cont.) & ENDIREH 2021\\
Age at cohabitation or marriage with current partner (cont.) & ENDIREH 2021\\
Consent to first sexual intercourse (cat.: no, yes) & ENDIREH 2021\\
Consent to current cohabitation or marriage (cat.: no, yes) & ENDIREH 2021\\
Pro-gender equality attitude (cat.: low, medium, high) & ENDIREH 2021\\
\hline
\multicolumn{2}{|l|}{\textbf{Relationship-Level}} \\
\hline
Age of partner in years (cont.) & ENDIREH 2021\\
Income per month of partner (cont.) & ENDIREH 2021\\
Violence witnessed in childhood by partner (cat.: no, yes)  & ENDIREH 2021\\
Violence experienced in childhood by partner (cat.: no, yes)  & ENDIREH 2021\\
Avg. household members per room (cont.) & ENDIREH 2021\\
Division of housework (cat.: both, only women, only men) & ENDIREH 2021\\
Woman's level autonomy about sex-life  (cat.: low, medium, high) & ENDIREH 2021\\
Woman's level autonomy about professional life and economic resources  (cat.: low, medium, high) & ENDIREH 2021\\
Woman's level autonomy about social and political activities  (cat.: low, medium, high) & ENDIREH 2021\\
Woman’s perception about support from social networks  (cat.: low, medium, high) & ENDIREH 2021\\
Level of social interaction  (cat.: low, medium, high) & ENDIREH 2021\\
\hline
\multicolumn{2}{|l|}{\textbf{Community-Level}} \\
\hline
Homicide rate of men per 100.000 men  (cont.) & Homicides 2021 \\
Homicide rate of women per 100.000 women  (cont.) & Homicides 2021 \\ 
Total homicide rate per 100.000 inhabitants  (cont.) & Homicides 2021 \\
Share of population in women-headed households  (cont.) & Intercensal 2020 \\
Female share of migrant population  (cont.) & Migration 2021 \\
Male share of migrant population  (cont.) & Migration 2021  \\
Gini index (cont.) & CONEVAL 2020 \\
Human development index  (cont.) & UNDP 2020 \\
Women's economically active population  (cont.) & Intercensal 2020 \\
Men's economically active population  (cont.) & Intercensal 2020 \\
Share of senior positions held by women  (cont.) & CNGMD 2020 \\
Level of social marginalization  (cat.: very low, low, medium, high, very high) & CONAPO 2020 \\
Type of community (cat.: rural, low urban, medium urban, high urban) & CONAPO 2020 \\
Municipality of residence & ENDIREH 2021\\
Centroid coordinates x and y of municipality & Geography Data \\
\hline
\multicolumn{2}{|l|}{\textbf{Societal-Level}} \\
\hline
Prevalence of common crimes against men per 100.000 men (cont.) & ENVIPE 2021 \\
Prevalence of common crimes against women per 100.000 women (cont.) & ENVIPE 2021 \\
Share of non-reported crimes against men  (cont.) & ENVIPE 2021 \\
Share of non-reported crimes against women  (cont.) & ENVIPE 2021 \\
Share of population that perceived corruption as a problem  (cont.) & ENCIG 2019 \\
Share of population satisfied with public services  (cont.) & ENCIG 2019 \\
Federal state of residence & ENDIREH 2021\\
\hline
 \end{longtable}
\end{center}

\subsection{Model Specification}
We employ an additive probit regression model to show the relationship between the factors and the likelihood of women experiencing psychological IPV within the past year. The outcome variable takes a value of 1 if the woman suffered from psychological IPV during the preceding year and 0 otherwise. This outcome variable follows a Bernoulli distribution denoted as $y_i \sim Bernoulli(\pi_i), i = 1,..., n$ with $\pi_i \in [0,1]$.

The probability of a woman $i$ suffering from psychological IPV between October 2020 and 2021 is modeled through the cumulative distribution function of the standard normal distribution applied to the additive predictor $\eta_i$. 
\begin{equation}
\begin{aligned}
  \pi_i &= \Phi(\eta_i) \\
       &= \Phi\Bigg(
          \beta_0 + 
          \sum_{p \in P}\beta_px_{ip} + \sum_{q \in Q}f_{q}(x_{iq})  + 
          \sum_{r \in R}f_r(\cdot) +  
          \sum_{s \in S}f_s(\cdot) \\
       &\quad + 
        f_{\text{spatial}}(x_{long,i},x_{lat,i})  +
       f_{random}(x_{state,i}) +
       f_{random}(x_{mun,i})  + 
       \epsilon_i
       \Bigg)
\end{aligned}
\label{model1}
\end{equation}
The specified model can be divided into the following parametric and non-parametric components. The intercept is represented by the term $\beta_0$. The model further incorporates linear effects for categorical variables. The set of categorical variables included in the model is denoted as $P$. Each factor level of these variables is treated as an individual effect, with one level selected as the reference category to establish a baseline for comparison. Continuous variables from the set $Q$ are centered around their mean and integrated as smooth effects. This integration is achieved using cubic P-Splines with second-order differences and 20 inner knots for the smooth function. The model employs P-spline decomposition to accommodate the choice between linear and smooth eff
cts for continuous variables. Furthermore, interaction effects\footnote{The interaction effects are: age difference between woman and partner * education level medium, age difference between woman and partner * education level high, age at first sexual intercourse * consent to sex, age at marriage * consent to marriage.} between continuous and categorical variables are incorporated. The set of continuous and categorical variables that interact with each other is represented by $R \varsubsetneq P \times Q$.  Consider an example interaction between a continuous variable $x$ and a categorical variable $z$. The interaction is captured by the function $f() = f_{\text{smooth}}(x) \cdot z$, where $f_{\text{smooth}}()$ is a smooth function representing the varying effect of $x$ in the presence of different levels of the categorical variable $z$. Again, P-spline decomposition is applied to distinguish whether there is no effect, a linear interaction effect or a non-linear effect between the variables. Plus, interaction surfaces\footnote{The interaction surfaces are: age at first child * age difference between woman and partner, age at first sexual intercourse * age difference between woman and partner, age at marriage * age difference between woman and partner, income woman * income partner.} $f_s(\cdot)$ between continuous variables and spatial effects related to the municipality's centroid longitude and latitude coordinates are estimated using bi-variate tensor product P-Splines with first-order differences and 20 inner knots. Additionally, random effects are incorporated to account for unobserved variability associated with states and municipalities. Hence, random intercepts are specified for each level of the factor variables. Finally, the variability that remains unaccounted for by the model is represented by the error term $\epsilon$.

\subsection{Analytical Strategy}
We follow a series of steps to estimate the specified model and identify the main risk and protective factors of psychological IPV. 

First, we use the model-based boosting algorithm to fit the specified generalized additive probit regression model in Eq. \ref{model1} \citep{friedmanAdditiveLogisticRegression2000, friedmanGreedyFunctionApproximation2001a, hothornMboostModelBasedBoosting2024a}. We adjust the model to the sample design and survey weights. In addition to model estimation, the algorithm performs automated (intrinsic) variable selection and model choice \citep{kneibVariableSelectionModel2009}. Thus, the algorithm identifies the most informative variables and chooses the appropriate modeling alternative, e.g., whether the variable is included with a linear, non-linear effect or not at all \citep{kneibVariableSelectionModel2009}. The boosting algorithm disposes of two notable tuning parameters: the shrinkage parameter $\nu$ and the number of boosting iterations $m_{stop}$. In our study, following \citet{torresmunguiaDeterminantsEmotionalIntimate2022}, we set the shrinkage parameter to $0.5$. Further, we tune the number of boosting iterations using the subsampling technique. Hence, we create 25 train sets, each containing 50\% randomly selected observations of the main data set. The iteration number that minimizes the average prediction error across all test sets is chosen as the optimal value for $m_{stop}$ \citep{mayrImportanceKnowingWhen2012}.

Second, we use the stability selection algorithm to refine the subset of variables selected by the boosting algorithm further \citep{meinshausenStabilitySelection2010, shahVariableSelectionError2013, hofnerControllingFalseDiscoveries2015}. Specifically, we generate 100 subsets, each comprising 50\% randomly selected observations of the main data set. We
fit boosting models to each of them. Variables selected in more than 80\% of the generated subsamples are considered significant and are included in the final set of stable variables. 

In the final step, we construct 95\% pointwise bootstrap confidence intervals for the effects of the selected variables. These are calculated from 1000 bootstrap replicates.

We use the following R-package: \textit{mboost} \citep{hofnerModelbasedBoostingHandson2014, hothornMboostModelBasedBoosting2024a} for all three analysis steps. The data and R-code to replicate the estimation are available in OSF at \url{https://osf.io/nefdj/?view_only=83803f190bf9442986a65dd2e18d3367}. In addition, we conducted two robustness checks to ensure the validity of our primary analysis (see Appendix \ref{app_rob_check}).

\section{Results}
\label{results}

In total, we identify 16 significant factors out of the 59 that were included in the model. All these variables are selected with a linear effect. Table \ref{maineffects} displays the results. Our analysis reveals six risk factors and ten protective factors among the selected variables. 

\begin{table}[h!]
\caption{Partial effects of the selected significant factors by ecological model level. For the selected effects, 95\% pointwise bootstrap confidence intervals have been included}
\label{maineffects}
\centering
\begin{tabular}{|p{0.6\textwidth}|p{0.1\textwidth}|l|}
\hline
\hline
\textbf{Factors} & \textbf{Estimate} & \textbf{95}\% \textbf{CI}\\
\hline
\textbf{Individual-Level }& &\\ \hline
Unemployment in the last 12 months yes  & -0.112 & [-0.138, -0.097]\\ 
Violence witnessed in childhood yes  & 0.235 & [0.209, 0.260] \\ 
Violence experienced in childhood yes  & 0.193 & [0.168, 0.220]\\ 
Sexual violence experienced in childhood yes & 0.311 & [0.276,    0.345]\\ 
Consent to sex at first sexual intercourse yes  & -0.3 & [-0.334, -0.249]\\ 
Women's age at first sexual intercourse by consent to sex yes  & -0.02 & see Figure \ref{part1} \\ 
Consent to current marriage or cohabitation yes  & -0.143& [-0.229  -0.132]\\  \hline
\textbf{Relationship-Level} & &\\ \hline
Violence witnessed by partner in childhood yes  &  0.246& [0.218,   0.273] \\
Violence experienced by partner in childhood yes  & 0.329& [0.300,   0.356]\\
Division of housework among male and female members  & -0.087& [-0.112, -0.050] \\
Division of housework only among male members  & -0.163& [-0.190, -0.124] \\
Woman’s level autonomy about sex-life medium  & -0.435& [-0.474,  -0.422]\\
Woman’s level autonomy about professional life and economic resources medium  & -0.357& [-0.418, -0.259] \\
Woman’s perception about support from social network high  & -0.092& [-0.125, -0.062]\\
Level of social interaction medium  & -0.058& [-0.079, -0.029]\\ \hline
\textbf{Societal-Level} & &\\ \hline
Share of non reported common crimes against men & 0.019& see Figure \ref{part2} \\  
\hline
\hline
\end{tabular}
\end{table}

\begin{figure}[h!]
  \centering
    \includegraphics[width=0.7\linewidth]{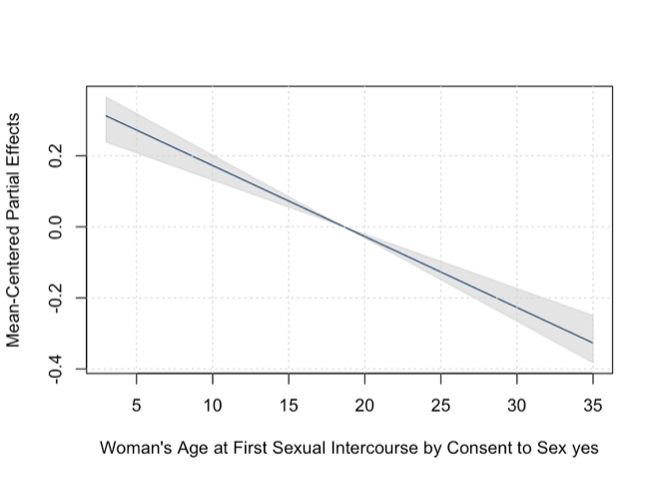}
      \caption{Partial effect of women's age at first sexual intercourse by consent to sex yes}
  \label{part1}
\end{figure}
\begin{figure}[h!]
  \centering
    \includegraphics[width=0.7\linewidth]{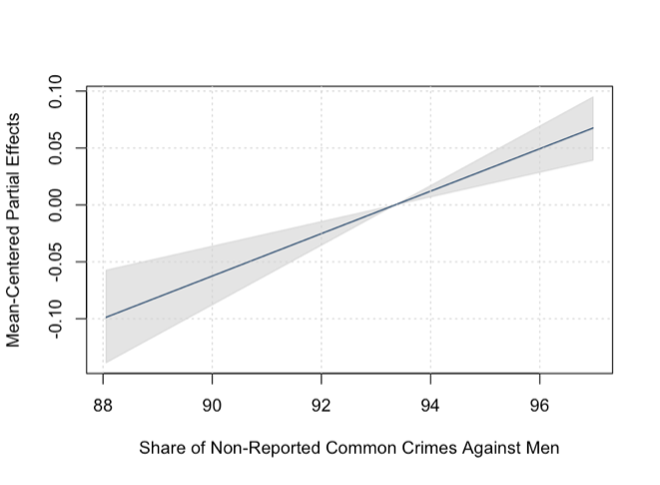}
      \caption{Partial effect for the share of non reported common crimes against men}
    \label{part2}
\end{figure}


\paragraph{Individual-Level.} 
At the individual level, we identify four protective and three risk factors. 

Our finding that women who were not employed in the last 12 months are at lower risk of suffering psychological IPV is consistent with existing research \citep{casiqueCambiosConstantesNiveles2019,rodriguez-hernandezPrevalenciaCorrelatosViolencia2023}. Women's financial independence and the shift in traditional gender roles might be threatening to their partners, who resort to violence to maintain control \citep{villarrealWomenEmploymentStatus2007a,bobonisPublicTransfersDomestic2013,canedoEstimationEffectWomen2021}. However, we cannot exclude a bi-directional relationship. Psychological IPV may affect women's ability to work and lead to unemployment \citep{guptaIntimatePartnerViolence2018}. Therefore, programs that aim to increase women's economic independence should be complemented by programs that address this risk \citep{eswaranDomesticViolenceWomen2011,terrazas-carrilloEmploymentStatusIntimate2015}; for example, by integrating social support services for women in the workplace \citep{canedoEstimationEffectWomen2021}.

Our other findings also contribute to and confirm existing literature showing an association between non-consensual early-life first-sex experiences and later-life negative physical and psychological health outcomes \citep{kaplanEarlyAgeFirst2013,mccarthy-jonesAssociationsForcedPersuaded2019}, as well as higher risk of experiencing psychological violence \citep{lalchandaniEarlyLifeSexual2020}. This result highlights the potential of addressing psychological IPV by introducing, for instance, sexuality education programs in schools. \citet{makleffPreventingIntimatePartner2020} also show the potential of such programs in preventing partner violence by reshaping gender and sexuality-related social norms in Mexico City. These education programs may be extended to teach consent \citep{burtonTeachingSexualConsent2023}. While our findings hold significant value, they should be interpreted with caution. Sexual consent in first-time experiences is a complex, multi-faceted concept that requires dedicated, in-depth research to be fully understood \citep{fennerSexualConsentScientific2017}. In our study, we assessed consent of first-sex experience by using the following two questions: ``How old were you when you had sex for the first time?'' followed by ``Was this first experience with your consent (you agreed to have sex)?'' \citep{inegiEncuestaNacionalSobre2021}. Hence, we rely on retrospective information, which raises three critical considerations. First, it challenges the extent to which a woman aged 15 or older can accurately recall the consensual nature of her first sexual experience \citep{willisMomentaryRetrospectiveSexual2022}. Second, it emphasizes the need to consider that perceptions of sexual consent may develop retrospectively and may differ from initial momentary perceptions \citep{willisMomentaryRetrospectiveSexual2022}. Third, it raises the concern of whether the question is a sufficient and valid measure of retrospective sexual consent or whether a measurement is possible at all \citep{willisDevelopingValidFeasible2021}.

Further, our results especially underline the importance of considering women's violence exposure in childhood when analyzing psychological IPV. We find that both experiencing or witnessing violence in childhood (referring to the period when the woman was under 15 years old) increases women's psychological IPV risk. Witnessing violence in childhood means being exposed to physical violence, verbal abuse, or insulting acts while living with others during that period. Experiencing such violence, on the other hand, signifies being a direct target of these actions. Notably, women who experienced violence in childhood face an approximately 19 percentage points higher psychological IPV risk than those who did not. Similarly, experiencing sexual violence during childhood, defined as coercion into sexual activities through force or threats during that period, increases the mothers' risk of becoming a psychological IPV victim by about 31 percentage points.
Our findings highlight two key aspects that should be considered by future research. First, childhood violence exposure should be considered when analyzing psychological IPV. Second, a differentiated analysis is necessary to understand why women who experience violence in childhood enter psychologically violent relationships and why men perpetuate this violence in their relationships. Overall, existing literature has studied and highlighted the intersection of violence against children (VAC) and IPV in different countries \citep{guedesBridgingGapsGlobal2016,herrenkohlChildMaltreatmentYouth2022,bottCorrelatesCooccurringPhysical2022}. Thereof, \citet{guedesBridgingGapsGlobal2016} have reviewed global literature and identified three key intersections. First, VAC and IPV share the same risk factors. Second, VAC and IPV co-occur within the same household. Third, VAC and IPV have an inter-generational impact. The last point is particularly relevant for our analysis. It highlights the existing link between childhood violence exposure and an increased risk of experiencing IPV or becoming an IPV perpetrator. Several theoretical approaches can explain the existing link thereof: individuals that were exposed to violence in childhood can be more accepting of violence as a problem-solving approach \citep{valdez-santiagoPrevalenceSeverityIntimate2013}, childhood violence exposure can disrupt stress regulation, impact mental health and lead to impulsive behaviors \citep{connorAggressionAntisocialBehavior2002,herrenkohlChildMaltreatmentYouth2022}. Further theories include attachment theory \citep{herrenkohlChildMaltreatmentYouth2022} and social learning \citep{doumasAdultAttachmentRisk2008}. Our study, as most studies, relies on cross-sectional data and is limited to correlational findings \citep{madiganTestingCycleMaltreatment2019}. Nonetheless, the findings suggest the potential of mitigating IPV by addressing VAC, for instance, through home visits and family therapy programs \citep{krugWorldReportViolence2002}.

\paragraph{Relationship-Level.}
At the relationship level, we identify six protective and two risk factors.

The results show that women living in households where tasks such as cleaning, cooking, and repairs are equally or mainly done by men have a lower risk of experiencing psychological IPV compared to women living in households where women do all the work. Further, we find that women with moderate decision-making power regarding their sexual life have a roughly 43 percentage point lower risk of psychological IPV compared to those with low autonomy in this area. Similarly, women with moderate autonomy concerning professional decisions and the use of economic resources experience a lower psychological IPV risk. 

A woman's strong social support network and a woman's moderate level of social interaction are negatively associated with psychological IPV risk. As suggested by \citet{richardsonSocialSupportIntimate2022}, a bi-directional and reinforcing relationship between IPV and social support may exist. On the one hand, social network support may reduce IPV by encouraging women, e.g., to cooperate in criminal prosecutions against their abusers \citep{goodmanObstaclesVictimsCooperation1999}. On the other hand, experiencing IPV can diminish social support, either through the perpetrator's actions limiting the woman's social interactions or the woman's fear of judgment or stigma \citep{richardsonSocialSupportIntimate2022}. Identifying the most impactful supporters in reducing IPV risk and understanding the mechanisms through which different types of support contribute to risk reduction are critically important. Hence, this finding highlights the urgent need to introduce social support programs and encourage women to use these resources. 

Additionally, our results indicate that increased social interactions of the woman, such as spending time with friends, talking to neighbors, or participating in team sports, are associated with a lower risk of IPV. Friends, neighbors, and team members may, in fact, act as informal support systems for women \citep{daviesSystematicReviewInformal2023}. A reciprocal and reinforcing dynamic may emerge: high levels of social interaction can be perceived as a threat by the partner, leading to psychological violence as a means of regaining control. In response, the woman may restrict her social interactions to protect herself from risks such as stalking or unfounded accusations. This two-way relationship must be considered when implementing programs and measures to strengthen women's social interaction. One specific aspect worth highlighting is women's participation in team sports. Research suggests that increased female sports involvement contributes to a lower risk of IPV, possibly because female athletes often challenge traditional gender norms \citep{milnerAthleticParticipationIntimate2017}.

We want to stress again the importance of childhood experiences when aiming to understand psychological IPV risk groups. We find that women are at higher psychological IPV risk when their partners have witnessed or experienced childhood violence. Women are almost 33 percentage points more likely to experience psychological IPV if their partner witnessed violence as a child, like observing their mother being abused.

\paragraph{Societal-Level.}
At the societal level, we identify one relevant risk factor.

We find that a higher prevalence of non-reported crimes against men, encompassing offenses like extortion and assault, at the federal-state level is associated with an increased risk of IPV. As shown by the EVIPE survey, two primary factors contribute to non-reporting: concerns about perceived time loss and a lack of trust in authorities \citep{inegiEncuestaNacionalVictimizacion2021}. A higher incidence of non-reporting can potentially lead to violent and abusive behavior, including in intimate relationships, as perpetrators may not fear reporting or legal consequences.

\section{Conclusion}
\label{conclusion}
In this study, we identify the most relevant risk and protective factors associated with psychological intimate partner violence (IPV) against women in Mexico in 2021. Following \citet{torresmunguiaDeterminantsEmotionalIntimate2022}, but using newer data sources, we also adopt an ecological model and study factors and their interactions at four levels: Individual, relationship, community, and societal. We use ten data sources to generate our data set. The 2021 Mexican Survey on the Dynamics of Household Relationships (ENDIREH 2021) serves as the cornerstone of our data set as it provides nationally representative information on women's experiences of psychological IPV in Mexico \citep{inegiEncuestaNacionalSobre2021}. In addition, the survey offers valuable data on socio-demographic characteristics and relationship dynamics, which are crucial for constructing individual and relationship-level factors. We integrate data from nine sources to include information on municipalities and states. This allows us to create community and societal-level factors. Our final constructed data set comprises 61,205 observations and 59 covariates. To identify the relationship between the factors and women's likelihood of experiencing psychological IPV, we use a machine learning algorithm. We find several factors (some of them newly included by us in the ecological model of psychological IPV) to be significant and relevant. Specifically, we identify a link between early and later life experiences of violence. The results indicate that women who experienced or witnessed psychological or physical violence in childhood, who suffered sexual violence in childhood, and whose partners were exposed to violence in childhood face a higher psychological IPV risk. Further, we find that employed women are more susceptible to psychological IPV compared to non-employed women. Additionally, the following results of our analysis are in line with and confirm the findings by \citet{torresmunguiaDeterminantsEmotionalIntimate2022}: Women with consensual first-sex experiences later in life face a lower risk of psychological IPV. Similarly, a lower risk is associated with women residing in households where men exclusively handle housework and have considerable decision-making power over their professional and economic lives. On the other hand, women living in states with high crime rates against men are at higher risk of experiencing psychological IPV.

We encourage future research to extend our study and suggest several theoretical and methodological extension possibilities. First, future research may construct an outcome variable that captures varying intensity levels of psychological IPV and evaluates women's risk of suffering different IPV levels. In our study, we relied on a binary measure of psychological IPV and set the threshold between suffering any or no psychological IPV. Second, future research should study different forms of IPV using newer data, for instance, physical, economic, and sexual IPV, as well as their co-occurrence \citep{palmer2024relationship}. Third, while we investigated psychological IPV risk for women with at least one child, further research can examine the risk for different groups of women, such as women without children, and identify potential differences in associated risk and protective factors. Furthermore, developing a causal model is essential to understanding the complex cause-and-effect relationships between various factors and psychological IPV. Applying methods such as propensity score matching can help estimate the impact of specific ecological model factors on the risk of experiencing psychological IPV. Our study relied on cross-sectional data and could not draw causal conclusions. Fourth, as the researcher determines the factors included at the four levels of the ecological model, there is room for feature engineering improvement. This involves adding more factors and improving the design of existing ones. Finally, studies may consider alternative approaches to the boosting algorithm and stability selection combination. For instance, use the method proposed by \citet{stromerDeselectionBaselearnersStatistical2022} to deselect variables based on their contribution to risk reduction within the boosting algorithm or use a generalized additive model in conjunction with various variable selection techniques.

To conclude, we highlight the urgency to understand better the mechanisms behind inter-generational transmission of violence in Mexico. A comprehensive understanding holds the potential to enhance programs and policies, ultimately preventing violence during childhood and mitigating its enduring repercussions on intimate partnerships.

\backmatter

\section*{Declarations}

\subsection*{Competing interests}
The authors declared no potential conflicts of interest with respect to the research, authorship, and/or publication of this article.

\subsection*{Funding}
The authors received no financial support for this article's research, authorship, and/or publication.

\subsection*{Ethics approval and consent to participate}
The data analyzed in this study were sourced from a publicly available open survey data set; therefore, no ethical approval was required.

\subsection*{Data availability}
The data and R-code to replicate the estimation are available in OSF at \url{https://osf.io/nefdj/?view_only=83803f190bf9442986a65dd2e18d3367}.

\subsection*{Consent for publication}
Not applicable.

\subsection*{Author contribution}
\begin{itemize}
    \item Conceptualization: Clara Strasser Ceballos and Anna-Carolina Haensch
    \item Data Analysis: Clara Strasser Ceballos
    \item Writing: Clara Strasser Ceballos
    \item Review and Editing: Anna-Carolina Haensch
\end{itemize}

\begin{appendices}

\section{Survey Data}
\label{app_survey_data}

\subsection{Sample Design}

The sample design of the Survey on the Dynamics of Household Relationships in Mexico (ENDIREH 2021) plays a crucial role in ensuring the validity of the sample and guaranteeing nationally and federally representative information \citep{inegiENDIREHDisenoMuestral2021}. The sample design encompasses the following characteristics: probabilistic, three-stage, stratified, and clustered sampling \citep{inegiENDIREHDisenoMuestral2021}.

The steps involved in the sampling process are the following \citep{inegiENDIREHDisenoMuestral2021}. The initial phase of the sample design entails the creation of Primary Sampling Units (PSUs). PSUs comprise a set of dwellings. The size of each PSU depends on the geographical area, i.e., highly urban, supplementary urban or rural. For instance, in a highly urban area, a PSU consists of a minimum of 80 and a maximum of 160 dwelling units. In the second phase, the PSUs are categorized into 683 different geographic and socio-demographic strata at the national level. The PSUs of the same stratum share common socio-demographic characteristics of the dwellers, physical characteristics of the dwellings, and geographical characteristics. In the third phase, the sampling selection is conducted in three stages. In the first stage, subsets of PSUs are selected within each geographical area with equal probability from each stratum and federal state. Within these selected PSUs, dwellings are chosen with equal probability to ensure unbiased representation. The number of dwellings selected in each PSU varies according to the geographical area. For instance, in a highly urban area, 5 dwellings are chosen with equal probability. Finally, within each chosen dwelling, a woman aged 15 years or older is randomly selected as a survey respondent for the ENDIREH 2021. This selected woman serves as the unit of observation in the survey data set. 

\subsection{Sample Size}

The sample size comprises 140,784 dwellings at the national level \citep{inegiENDIREHDisenoMuestral2021}. These dwellings are equally distributed across the 32 federal states of Mexico. Information about women's experiences of violence was collected for 110,127 women aged 15 and over who were permanent residents of the selected dwellings. Surveys were conducted over one month, from the 4th of October to the 30th of November 2021. The information collected can be disaggregated geographically at two levels: a) national - urban and rural, and b) by federal state.

\subsection{Questionnaire Structure}

\begin{figure}[h!]
  \begin{center}     \includegraphics[width=0.8\textwidth,height=0.4\textheight,keepaspectratio]{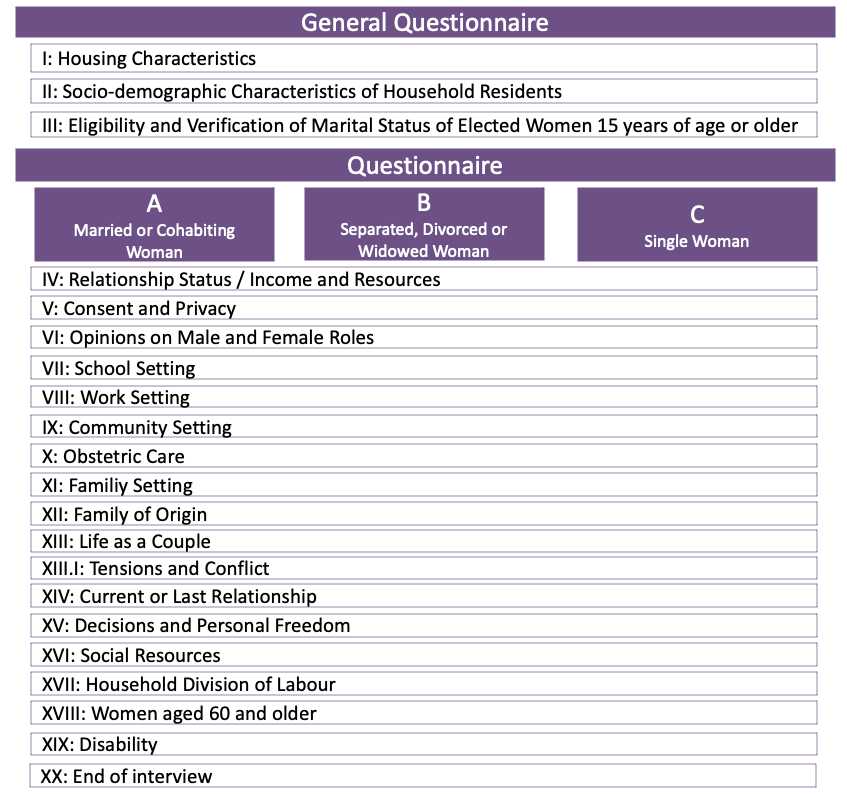}
  \end{center}
        \caption{Questionnaire structure}
    \label{quest_struc}%
\end{figure}

Four instruments were used to collect information on the 140,784 selected dwellings and the 110,127 women who participated in the study \citep{inegiENDIREHInformeOperativo2021}.  The general questionnaire is the central instrument. The questionnaire is divided into three sections. The first section collects data on household composition (I), the second section collects socio-demographic characteristics of all household members (II), and the third section identifies all women aged 15 years and above in the household and verifies their marital status (III). The three marital statuses are: married or cohabiting (A), separated, divorced or widowed (B), and single (C). An appropriate informant completes this questionnaire. Hence, a person aged 15 or over who lives in the dwelling has extensive knowledge of all other residents. Hereafter, one woman is selected as the household's observation unit. Depending on her marital status, she is assigned one of three specific questionnaires: A, B, or C. These three questionnaires comprise 20 thematic sections (IV-XX) covering various topics. An overview of the topics covered in both the general questionnaire and questionnaires A, B, and C can be found in Figure \ref{quest_struc}. The questionnaires were completed electronically using a mobile device called Meebox.

\section{Data Preparation}
\label{app_data_prep}

In this study, we used the ENDIREH 2021 data set as the baseline for generating our data set. The ENDIREH 2021 data set includes information on 110,127 women across Mexico. Our data preparation involved the following steps.

\paragraph{Step 1}
We filtered the ENDIREH 2021 data set to include only women who were 15 years or older, married or cohabiting with a male partner, and had at least one child. We created variables from the questions in the survey data set and integrated additional variables related to municipality and state-level factors. The filtered data set contained 63,152 observations covering over 80 factors.

\paragraph{Step 2}
To maintain the data set's integrity and avoid information loss, we employed imputation techniques, specifically predictive mean matching and single data imputation, to impute missing values \citep{buurenFlexibleImputationMissing2018}. We utilized the R-package \textit{mice} \citep{buurenMiceMultivariateImputation2023} to execute this imputation process. 

\paragraph{Step 3}
Once we imputed the missing values and assessed the imputation for correctness, we conducted a plausibility analysis. We performed a verification test to ensure that none of the following situations occurred and removed any observations from the data set where these conditions were met:
\begin{itemize}
    \item Age at first sexual intercourse of woman may not be $>$ than the age of woman.
    \item Age at first marriage or cohabitation of woman may not be $>$ than the age of woman.
    \item Age at first childbirth may not be $>$ than the age of woman.
    \item Age at first sexual intercourse of woman may not be $>$ than the age at first childbirth.
    \item Income of woman may not be $>$ 0, if the woman was never employed in her life.
\end{itemize}

We removed 378 observations in total. These removals were primarily due to two implausible situations:

\begin{enumerate}
    \item Many women reported having an income despite not being employed. The income question refers explicitly to employment-based income.
    \item Many women reported an age of first sexual intercourse that was higher than their age at first childbirth. This gap may be due to a misunderstanding of the question ``How old were you when you had sex for the first time?'' which is asked in the ``Couple Life'' Section of the Questionnaire A \citep{inegiENDIREHInformeOperativo2021}. Prior to this question, women were asked about their partner, which might lead them to think the question refers to their first sexual intercourse with their current partner or husband.
\end{enumerate}

\paragraph{Step 4}
Finally, we removed outliers from the 62,774 observations identified in the third step. These outliers were detected using box plots. After this process, a total of 61,205 observations remained for the main analysis.

\section{Robustness Checks}
\label{app_rob_check}

We conducted two robustness checks to ensure the validity of the primary analysis results.

First, we estimated a generalized linear probit regression model, incorporating all relevant risk and protective factors identified by the stability selection algorithm in the main study, we included sampling weights and an offset in the model. This extended analysis aimed to confirm that the effects identified by the model-based boosting algorithm were consistent with those of a generalized linear model. The results, displayed in Table \ref{rob1}, validate the effects identified by the model-based boosting algorithm. All factors exhibit similar effect sizes as those found in the primary analysis (see Section \ref{results}). Notably, the factor ``woman's age at first sexual intercourse,'' which was not selected by the stability selection algorithm, does not appear to be statistically significant. It only shows a negative association with the risk of psychological IPV when interacted with the variable ``consent to first sexual intercourse.'' Using a generalized linear model provides standard errors for the results and significance levels, thereby enhancing the overall robustness of the analysis.

\begin{table}[ht!]
\caption{Generalized linear model results}
\label{rob1}
\begin{tabular}{p{0.5\textwidth}|p{0.09\textwidth}|p{0.08\textwidth}|p{0.08\textwidth}|p{0.08\textwidth}}
  \hline
  \hline
\textbf{Factors} & \textbf{Estimate} & \textbf{Std. Error} & \textbf{t value} & \textbf{Pr($>$$|$t$|$)} \\ 
  \hline
  \textbf{Individual-Level} & & & &\\ \hline
  Intercept & -0.1901 & 0.0928 & -2.05 & 0.0406 \\ 
  Unemployment in the last 12 months yes & -0.1361 & 0.0177 & -7.71 & 0.0000 \\ 
  Violence witnessed in childhood yes & 0.2399 & 0.0198 & 12.13 & 0.0000 \\ 
  Violence experienced in childhood yes & 0.1976 & 0.0200 & 9.90 & 0.0000 \\ 
  Sexual violence experienced in childhood yes & 0.3688 & 0.0275 & 13.42 & 0.0000 \\ 
  Consent to sex at first sexual intercourse yes & -0.2407 & 0.0757 & -3.18 & 0.0015 \\ 
  Women's age at first sexual intercourse & 0.0237 & 0.0107 & 2.23 & 0.0261 \\ 
  Women's age at first sexual intercourse by consent to sex yes & -0.0450 & 0.0110 & -4.10 & 0.0000 \\ 
   \hline
\textbf{Relationship-Level} & & & &\\ \hline
  Consent to current marriage or cohabitation yes & -0.2009 & 0.0551 & -3.64 & 0.0003 \\ 
  Violence witnessed in childhood by partner yes & 0.2474 & 0.0212 & 11.66 & 0.0000 \\ 
  Violence experienced in childhood by partner yes & 0.3332 & 0.0208 & 16.04 & 0.0000 \\ 
  Division of housework among both & -0.1217 & 0.0227 & -5.36 & 0.0000 \\ 
  Division of housework among men& -0.2040 & 0.0231 & -8.82 & 0.0000 \\ 
  Woman’s level autonomy about sex-life medium & -0.4271 & 0.0234 & -18.22 & 0.0000 \\ 
  Woman’s level autonomy about professional life and economic resources medium & -0.2356 & 0.0197 & -11.96 & 0.0000 \\ 
  Woman’s perception about support from social network high & -0.0955 & 0.0240 & -3.98 & 0.0001 \\ 
  Level of social interaction medium & -0.0777 & 0.0186 & -4.18 & 0.0000 \\ 
  \hline
  \textbf{Societal-Level} & & & & \\ \hline
  Share of non reported common crimes against men  & 0.0319 & 0.0043 & 7.42 & 0.0000 \\ 
   \hline
   \hline
\end{tabular}
\end{table}

As a second robustness test, we applied the boosting algorithm to a data set comprising only complete cases, with no imputed missing data. After re-executing the data preparation steps, the data set was reduced to 26,889 observations. The stability selection algorithm identified 15 relevant variables out of the 59 initially specified. The results of Table \ref{rob2} validate and reinforce the main findings obtained from imputed data. Except for non-reported crimes against men, the stability selection algorithm consistently identifies the same set of risk and protective factors as in the main analysis. This robustness test underscores the significance of childhood violence exposure and underscores the importance of incorporating childhood experiences into programs aimed at addressing IPV.

\begin{table}[ht!]
\caption{Partial effects of the selected significant factors by ecological model level. For the selected effects, 95\% pointwise bootstrap confidence intervals have been included}
\label{rob2}
\centering
\begin{tabular}{|p{0.6\textwidth}|p{0.1\textwidth}|l|}
\hline
\hline
\textbf{Factors} & \textbf{Estimate} & \textbf{95}\% \textbf{CI}\\
\hline
\textbf{Individual-Level} & & \\ \hline
Unemployment in the last 12 months yes & -0.139 & [-0.167, -0.101] \\
Violence witnessed in childhood yes & 0.226 & [ 0.182, 0.259]\\
Violence experienced in childhood yes & 0.218 & [0.176, 0.250]\\
Sexual violence experienced in childhood yes & 0.273 & [0.220, 0.314]\\
Consent to sex at first sexual intercourse yes & -0.172 & [-0.264, -0.087]\\
Women's age at first sexual intercourse by consent to sex yes & -0.017 & Figure \ref{part_effects3}\\
Consent to current marriage or cohabitation & -0.241 & [-0.324, -0.127]\\ \hline
\textbf{Relationship-Level} & & \\ \hline
Violence witnessed in childhood by partner yes & 0.236 & [0.190, 0.274]\\
Violence experienced in childhood by partner yes & 0.323 & [0.280, 0.361]\\
Division of housework among both & -0.117 & [ -0.165, -0.070]\\
Division of housework among men & -0.227 & [-0.279, -0.179]\\
Woman’s level autonomy about sex-life medium & -0.428 & [-0.483, -0.383]\\
Woman’s level autonomy about professional life and economic resources medium & -0.348 & [-0.427, -0.238]\\
Woman’s perception about support from social network high & -0.120 & [-0.162, -0.065]\\
Level of social interaction medium & -0.072 & [-0.110, -0.035]\\ \hline
\hline
\hline
\end{tabular}
\end{table}

\begin{figure}[h!]
  \centering
    \includegraphics[width=0.7\linewidth]{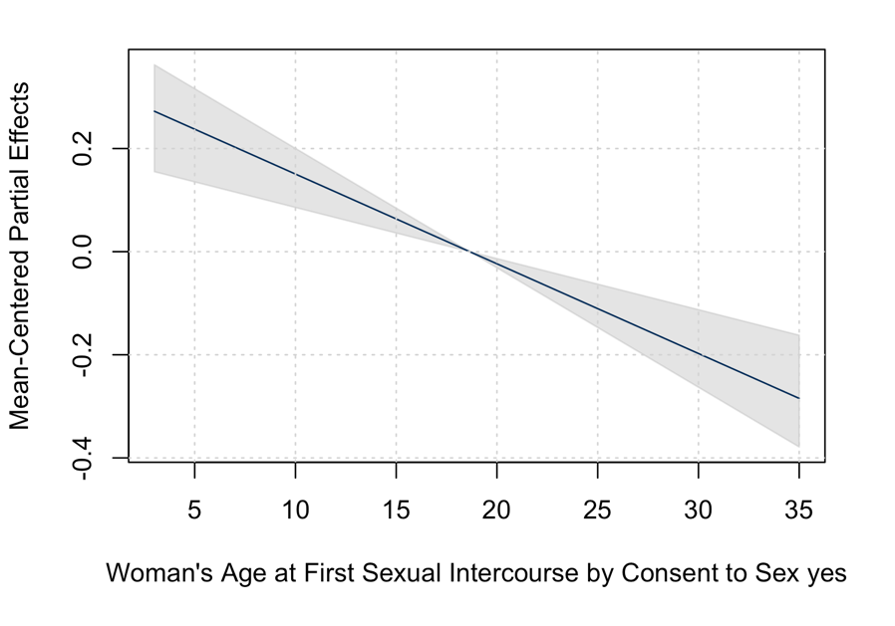}
      \caption{Partial effect of woman's age at first sexual intercourse by consent to sex yes}
  \label{part_effects3}
\end{figure}

\newpage
\section{Variable Codebook}
\label{app_code_book}

The following codebook provides information about the created variables, including their meaning, type, name in the data set and source. This codebook is an extended version of the codebook by the authors \citep{torresmunguiaDeterminantsEmotionalIntimate2022}.

    \begin{longtable}{|p{0.3\textwidth}|p{0.7\textwidth}|}
\caption{Codebook}
\label{tab_code} 
     \\ \hline
    \textbf{Variable} & \textbf{Description} \\
    \hline
    \endfirsthead
    \multicolumn{2}{l}{\textit{}} \\
    \hline
    \textbf{Variable} & \textbf{Description} \\
    \hline
    \endhead
    \hline
    \multicolumn{2}{r}{\textit{}} \\
    \endfoot
    \hline
    \endlastfoot
\multicolumn{2}{|l|}{\textbf{Individual-Level}} \\
\hline
Woman's age & Age of woman in years\\
& \textbf{Type:} Continuous\\
& \textbf{Name in data set:} \textit{EDAD}\\
& \textbf{Source:} \citet{inegiEncuestaNacionalSobre2021}\\
\hline
Woman's income & Woman's reported monthly income in Mexican Pesos\\
& \textbf{Type:} Continuous\\
& \textbf{Name in data set:} \textit{ingm\_muj}\\
& \textbf{Source:} \citet{inegiEncuestaNacionalSobre2021}\\
\hline
Woman's age at first childbirth & Age in years of the woman at first childbirth\\
& \textbf{Type:} Continuous\\
& \textbf{Name in data set:} \textit{eda\_hij}\\
& \textbf{Source:} \citet{inegiEncuestaNacionalSobre2021}\\
\hline
Woman's age at first sexual intercourse & Age in years of the woman at her first sexual intercourse\\
& \textbf{Type:} Continuous\\
& \textbf{Name in data set:} \textit{eda\_sex}\\
& \textbf{Source:}  \citet{inegiEncuestaNacionalSobre2021}\\
\hline
Indigenous origin & Indigenous self-identification of the woman\\
& \textbf{Type:} Categorical\\
& -"yes": If the woman self identifies as an indigenous \\
& -"no": Otherwise\\
& \textbf{Name in data set:} \textit{indigena}\\
& \textbf{Source:} \citet{inegiEncuestaNacionalSobre2021}\\
\hline
Education level & Degree of formal education level completed by the woman\\
& \textbf{Type:} Categorical\\
& -"low": None, preschool, primary education\\
& -"medium": At least secondary education and maximum high school\\
& -"high": At least university\\
& \textbf{Name in data set:} \textit{niv\_ed}\\
& \textbf{Source:} \citet{inegiEncuestaNacionalSobre2021} \\
\hline
Woman's consent to first sexual intercourse & Consent to first sexual intercourse\\
& \textbf{Type:} Categorical\\
& -"yes": If the woman consented to her first sexual intercourse\\
& -"no": Otherwise\\
& \textbf{Name in data set} \textit{con\_sex}\\
& \textbf{Source:} \citet{inegiEncuestaNacionalSobre2021}\\
\hline
Pro-gender equality attitude & Opinion on gender roles in childcare, income, employment, responsibility for housework, women's right to go out at night, whether men are more able to work and study, whether women with children who work neglect their responsibilities as mothers, whether women should have sex whenever their partner wants them to, the kind of clothes women should wear\\
& \textbf{Type:} Categorical\\
& -"low": If the woman has a pro-gender equality opinion in 2 to less questions\\
& -"medium": If the woman has a pro-gender equality opinion in more than 2 but less than 6 questions\\
& -"high": If the woman has a pro-gender equality opinion in more than 6 questions\\
& \textbf{Name in data set:} \textit{feminist\_grad}\\
& \textbf{Source:} \citet{inegiEncuestaNacionalSobre2021}\\
\hline
CCT receiver & Woman receives conditional cash transfers (CCT) like Prospera\\
& \textbf{Type:} Categorical\\
& -"yes": If the woman receives CCT\\
& -"no": Otherwise\\
& \textbf{Name in data set:} \textit{cct\_rec}\\
& \textbf{Source:} \citet{inegiEncuestaNacionalSobre2021}\\
\hline
Woman's unemployment & Woman was unemployed in the last 12 months\\
& \textbf{Type:} Categorical\\
& -"yes": If the woman did not have a paid employment in the last 12 months\\
& -"no": Otherwise\\
& \textbf{Name in data set:} \textit{desempleo}\\
& \textbf{Source:} \citet{inegiEncuestaNacionalSobre2021}\\
\hline
Previous marriage or cohabitation & If the woman had a previous cohabitation or marriage before her current partner or husband\\
& \textbf{Type:} Categorical\\
& -"yes": If the woman has had more than one partner\\
& -"no": Otherwise\\
& \textbf{Name in data set:} \textit{pareja\_prev}\\
& \textbf{Source:} \citet{inegiEncuestaNacionalSobre2021}\\
\hline
Violence witnessed in childhood by woman & If the woman witnessed beating or insults in the household she lived when she was younger than 15 years old\\
& \textbf{Type:} Categorical\\
& -"yes": If the woman witnessed either beating or insults\\
& -"no": Otherwise\\
& \textbf{Name in data set:} \textit{vio\_inf}\\
& \textbf{Source:} \citet{inegiEncuestaNacionalSobre2021}\\
\hline
Violence experienced in childhood by woman & If the woman was beaten or insulted by a person in the household where she lived when she was younger than 15 years old\\
& \textbf{Type:} Categorical\\
& -"yes": If the woman experienced either beaten or insulted\\
& -"no": Otherwise\\
& \textbf{Name in data set:} \textit{vio\_exp\_inf}\\
& \textbf{Source:} \citet{inegiEncuestaNacionalSobre2021}\\
\hline
Sexual violence experienced in childhood by woman & If the woman was sexually abused (forced to have sex, forced to perform sexual acts, was forced to show her private parts, was touched by someone,...) by a person in the household where she lived when she was younger than 15 years old\\
& \textbf{Type:} Categorical\\
& -"yes": If the woman was sexually abused\\
& -"no": Otherwise\\
& \textbf{Name in data set:} \textit{vio\_sex\_inf}\\
& \textbf{Source:} \citet{inegiEncuestaNacionalSobre2021}\\
\hline
\multicolumn{2}{|l|}{Relationship-Level}\\
\hline
Woman's age at marriage or at cohabitation & Age in years of the woman at marriage with current husband or at cohabiting with current partner\\
& \textbf{Type:} Continuous\\
& \textbf{Name in data set:} \textit{eda\_mat}\\
& \textbf{Source:} \citet{inegiEncuestaNacionalSobre2021}\\
\hline
Partner's age & Age in years of the partner\\
& \textbf{Type:} Continuous\\
& \textbf{Name in data set:} \textit{eda\_par2}\\
& \textbf{Source:} \citet{inegiEncuestaNacionalSobre2021}\\
\hline
Partner's income & Partner's reported monthly income in Mexican Pesos\\
& \textbf{Type:} Continuous\\
& \textbf{Name in data set:} \textit{ingm\_par}\\
& \textbf{Source:} \citet{inegiEncuestaNacionalSobre2021}\\
\hline
Violence witnessed in childhood by partner & If the husband or partner witnessed beating or insults in the household he lived when he was younger than 15 years old\\
& -"yes": If the partner witnessed beating or insults\\
& -"no": Otherwise\\
& \textbf{Name in data set:} \textit{vio\_inf\_par}\\
& \textbf{Source:} \citet{inegiEncuestaNacionalSobre2021}\\
\hline
Violence experienced in childhood by partner & If the husband or partner was beaten or insulted by a person in the household he lived when he was younger than 15 years old\\
& -"yes": If the partner was beaten or insulted\\
& -"no": Otherwise\\
& \textbf{Name in data set:} \textit{vio\_exp\_inf\_par}\\
& \textbf{Source:} \citet{inegiEncuestaNacionalSobre2021}\\
\hline
Overcrowding & Average number of household members per room in the dwelling\\
& \textbf{Type:} Continuous\\
& \textbf{Name in data set:} \textit{hacin}\\
& \textbf{Source:} \citet{inegiEncuestaNacionalSobre2021}\\
\hline
Woman's consent to current marriage or cohabitation & Consent to marriage with current husband or for cohabiting with current partner\\
& \textbf{Type:} Categorical\\
& -"yes": If the woman consented to marriage or cohabitation\\
& -"no": Otherwise\\
& \textbf{Name in data set:} \textit{con\_mat}\\
& \textbf{Source:} \citet{inegiEncuestaNacionalSobre2021}\\
\hline
Division of housework among household members & Division of housework among household members\\
& \textbf{Type:} Categorical\\
& -"only woman": If housework at home is carried out only by female members\\
& -"both": If both women and men take part in housework\\
& -"only men": If housework is carried out only by male members\\
& \textbf{Name in data set:} \textit{act\_dist}\\
& \textbf{Source:} \citet{inegiEncuestaNacionalSobre2021}\\
\hline
Woman's level of autonomy within the relationship to make decisions about her sexual life &  Woman's perception of her level of freedom and autonomy in the relationship to make decisions about her sex life (e.g. when to have sex, who should use contraceptives, when to have children)\\
& \textbf{Type:} Categorical\\
& -"low": If woman expressed having non or less decision making power than her husband or partner\\
& -"medium": If the woman expressed having the same decision making power than her husband or partner\\
& -"high": If the woman expressed having all or more decision making power than the husband or partner\\
& \textbf{Name in data set:} \textit{lib\_sex\_grad}\\
& \textbf{Source:} \citet{inegiEncuestaNacionalSobre2021}\\
\hline
Woman's level of autonomy within the relationship to make decisions about her professional life and use of economic resources & The woman's perception of her level of freedom and autonomy within the relationship to make decisions about her professional life and use of economic resources (e.g. whether she can work or study, what she can do with the money she earns, whether she can buy things for herself)\\
& \textbf{Type:} Categorical\\
& -"low": If the woman expressed having none or less decision making power than her husband or partner\\
& -"medium": If the woman expressed having the same decision making power than her husband or partner\\
& -"high": If the woman expressed having all or more decision making power than her husband or partner\\
& \textbf{Name in data set:} \textit{lib\_eco\_grad}\\
& \textbf{Source:} \citet{inegiEncuestaNacionalSobre2021} \\
\hline
Woman's level of autonomy within the relationship to make decisions about her participation in social and political activities & The woman's perception of her level of freedom and autonomy within the relationship to make decisions about her participation in social and political activities (e.g. whether she can go out alone, whether she can participate in the social or political life of her community, whether she can decide what to wear when she goes out)\\
& \textbf{Type:} Categorical\\
& -"low": If the woman expressed having none or less decision making power than her husband or partner\\
& -"medium": If the woman expressed having the same decision making power than her husband or partner\\
& -"high": If the woman expressed having all or more decision making power than her husband or partner\\
& \textbf{Name in data set:} \textit{lib soc\_grad}\\
& \textbf{Source:} \citet{inegiEncuestaNacionalSobre2021} \\
\hline
Woman's perception of social support networks & The woman's perception of whether she could get support from social networks in some hypothetical situations, e.g. if the woman needs help with childcare, performing a task or work, if she is ill, if she needs to talk about her problems and worries, if she needs advice or help with problems with her husband or partner, and if she needs support in difficult economic times\\
& \textbf{Type:} Categorical\\
& -"low": If the woman considers she could get support from social networks in 1 or less situations\\
& -"medium": If the woman considers she could get support from social networks in more than 1 but less than 5 situations\\
& -"high": If the woman considers she could get support from social networks in at least 5 situations.\\
& \textbf{Name in data set:} \textit{redsoc\_grad}\\
& \textbf{Source:} \citet{inegiEncuestaNacionalSobre2021} \\
\hline
Woman's level of social interaction & Level of social interaction reported by the woman (e.g. going out with friends, talking with neighbors, meeting with family members, attending religious events, participating in organizations, and practicing team sports)\\
& \textbf{Type:} Categorical\\
& -"low": If the woman states she carries out 1 or less situations\\
& -"medium": If the woman states she carries out at least more than 1 but less than 5 situations\\
& -"high": If the woman states she carries out at least 5 situations.\\
& \textbf{Name in data set:} \textit{rout\_grad}\\
& \textbf{Source:} \citet{inegiEncuestaNacionalSobre2021} \\
\hline
\multicolumn{2}{|l|}{Community-Level} \\
\hline
Level of social marginalization & Level of social marginalisation in 2020 in the municipality of residence. This indicator takes into account nine socio-economic indicators at the community level: percentage of the population aged 15 and over who are illiterate; percentage of the population aged 15 and over who have not completed primary school; percentage of the population living in dwellings without drains and toilets; percentage of the population living in dwellings without electricity; percentage of the population living in dwellings without piped water; percentage of the population living in overcrowded dwellings (number of household members per room, including kitchen but excluding corridors and bathrooms, is greater than 2. 5); percentage of the population living in dwellings with dirty floors; percentage of the population living in settlements with less than 5,000 inhabitants; and percentage of the working population with an income of up to two minimum wages.\\
& \textbf{Type:} Categorical\\
& -"very low"\\
& -"low"\\
& -"medium "\\
& -"high"\\
& -"very high"\\
& \textbf{Name in data set:} \textit{Marg20}\\
& \textbf{Source:} \citet{conapoIndicesMarginacion2020} \\
\hline
Type of community & Type of community of the municipality of residence according to their number of inhabitants\\
& \textbf{Type:} Categorical\\
& -"rural": If the community has less than 2,500 inhabitants\\
& -"low urban": If the community has between 2,500 and 14,999 inhabitants\\
& -"medium urban": If the community has between 15,000 and 99,999 inhabitants\\
& -"high urban": If the community has 100,000 inhabitants or more inhabitants.\\
& \textbf{Name in data set:} \textit{Type\_com}\\
& \textbf{Source:} \citet{conapoIndicesMarginacion2020} \\
\hline
Homicide rate of women & Homicide rate of women from 2017 to 2021 divided by 2020 population in the municipality of residence multiplied by 100,000\\
& \textbf{Type:} Continuous\\
& \textbf{Name in data set:} \textit{fhr20}\\
& \textbf{Source:} \citet{inegiMortalidad2021} \\
\hline
Homicide rate of men & Homicide rate of men from 2017 to 2021 divided by 2020 population in the municipality of residence multiplied by 100,000\\
& \textbf{Type:} Continuous\\
& \textbf{Name in data set:} \textit{mhr20}\\
& \textbf{Source:} \citet{inegiMortalidad2021} \\
\hline
Total homicide rate & Total homicide rate from 2017 to 2021 divided by 2020 population in the municipality of residence multiplied by 100,000\\
& \textbf{Type:} Continuous\\
& \textbf{Name in data set:} \textit{ghr20}\\
& \textbf{Source:} \citet{inegiMortalidad2021} \\
\hline
Women's household headship & Share of population in women-headed households in the municipality of residence 2020 \\
& \textbf{Type:} Continuous\\
& \textbf{Name in data set:} \textit{phogief\_f}\\
& \textbf{Source:} \citet{conapoIndicesMarginacion2020} \\
\hline
Migration of women & Percentage of foreign-born female population residing in the municipality of residence in 2020 \\
& \textbf{Type:} Continuous\\
& \textbf{Name in data set:} \textit{pres2020\_f}\\
& \textbf{Source:} \citet{inegiMigracion2021} \\
\hline
Migration of men & Percentage of foreign-born male population residing in the municipality of residence in 2020 \\
& \textbf{Type:} Continuous\\
& \textbf{Name in data set:} \textit{pres2020\_m}\\
& \textbf{Source:} \citet{inegiMigracion2021}  \\
\hline
Gini index & Gini index in 2020 of the municipality of residence\\
& \textbf{Type:} Continuous\\
& \textbf{Name in data set:} \textit{gini20}\\
& \textbf{Source:} \citet{conevalMedicionPobreza2020} \\
\hline
Human development index & Human development index in 2020 of the municipality of residence\\
& \textbf{Type:} Continuous\\
& \textbf{Name in data set:} \textit{idh2020}\\
& \textbf{Source:} \citet{undpHumanDevelopmentIndex2020} \\
\hline
Women's economically active population & Percentage of economically active female population aged 12 and over in the municipality of residence in 2020\\
& \textbf{Type:} Continuous\\
& \textbf{Name in data set:} \textit{pea\_f}\\
& \textbf{Source:} \citet{inegiCensoPoblacionVivienda2020} \\
\hline
Men's economically active population &  Percentage of economically active male population aged 12 and over in the municipality of residence in 2020 \\
& \textbf{Type:} Continuous\\
& \textbf{Name in data set:} \textit{pea\_m}\\
& \textbf{Source:} \citet{inegiCensoPoblacionVivienda2020}  \\
\hline
Woman's political participation & Share of senior positions in the local public administration held by women in 2020 in the municipality of residence\\
& \textbf{Type:} Continuous \\
& \textbf{Name in data set:} \textit{ParPolF}\\
& \textbf{Source:} \citet{inegiCensoNacionalGobiernos2020} \\
\hline
\multicolumn{2}{|l|}{Societal-Level} \\
\hline
Common crimes against women & Prevalence rate of common crimes against women aged 18 or more per 100,000 women in 2020\\
& \textbf{Type:} Continuous\\
& \textbf{Name in data set:} \textit{FemPrev}\\
& \textbf{Source:} \citet{inegiEncuestaNacionalVictimizacion2021} \\
\hline
Common crimes against men & Prevalence rate of common crimes against men aged 18 or more per 100,000 men in 2020 \\
& \textbf{Type:} Continuous\\
& \textbf{Name in data set:} \textit{MasPrev}\\
& \textbf{Source:} \citet{inegiEncuestaNacionalVictimizacion2021} \\
\hline
Dark figure of common crimes against women & Share of common crimes against women aged 18 or more not reported to or not registered by the authorities in 2020\\
& \textbf{Type:} Continuous\\
& \textbf{Name in data set:} \textit{FemNoDen}\\
& \textbf{Source:} \citet{inegiEncuestaNacionalVictimizacion2021} \\
\hline
Dark figure of common crimes against men & -Share of common crimes against men aged 18 or more not reported to or not registered by the authorities in 2020 \\
& \textbf{Type:} Continuous\\
& \textbf{Name in data set:} \textit{MasNoDen}\\
& \textbf{Source:} \citet{inegiEncuestaNacionalVictimizacion2021} \\
\hline
Corruption & Share of 2019 population aged 18 years and over who considered corruption as a common or very common problem in their federal state of residence \\
& \textbf{Type:} Continuous\\
& \textbf{Name in data set:} \textit{cor19}\\
& \textbf{Source:} \citet{inegiEncuestaNacionalCalidad2019} \\
\hline
Satisfaction with public services & Share of 2019 population aged 18 years and over who were satisfied with the basic and on-demand public services provided in their federal state of residence\\
& \textbf{Type:} Continuous\\
& \textbf{Name in data set:} \textit{satis19}\\
& \textbf{Source:} \citet{inegiEncuestaNacionalCalidad2019} \\
\hline

\end{longtable}

\end{appendices}


\bibliography{mexico_ipv}

\end{document}